\documentclass[aps,pra,10pt,english,showpacs,floatfix,twocolumn,twoside]{revtex4-1}
\usepackage{amsmath,amstext,amssymb,babel,graphicx,geometry,bm,dcolumn,color,times,soul, braket, enumerate}
\usepackage[colorlinks=true,citecolor=blue,urlcolor=green]{hyperref}
\usepackage[table]{xcolor}
\begin{document}

\title{Scale-invariant freezing of entanglement}
\author{Titas Chanda$^{1}$, Tamoghna Das$^{1}$, Debasis Sadhukhan$^{1}$, Amit Kumar Pal$^{1,2}$, Aditi Sen(De)$^{1}$, Ujjwal Sen$^{1}$}
\affiliation{$^1$Harish-Chandra Research Institute, HBNI, Chhatnag Road, Jhunsi, Allahabad 211019, India \\
$^2$Department of Physics, Swansea University, Singleton Park, Swansea SA2 8PP, United Kingdom}

\begin{abstract}
We show that bipartite entanglement in a one-dimensional quantum spin model undergoing time-evolution under local Markovian environments can be \emph{frozen} over time. We demonstrate this by using a number of paradigmatic quantum spin models in one dimension, including the anisotropic XY model in the presence of a uniform and an alternating transverse magnetic field (ATXY), the XXZ model,  the XYZ model, and the $J_1-J_2$ model involving the next-nearest-neighbor interactions. We  show that  the length of the freezing interval, for a chosen pair of nearest-neighbor spins, may remain  independent of the length of the spin-chain, for example,  in paramagnetic phases of the ATXY model, indicating a scale-invariance. Such freezing of entanglement is found to be robust against a change in the environment temperature, presence of disorder in the system, and whether the noise is dissipative, or not dissipative. Moreover, we connect the freezing of entanglement with the propagation of information through a quantum many-body system, as considered in the Lieb-Robinson theorem. We demonstrate that the variation of the freezing duration exhibits a quadratic behavior against the distance of the nearest-neighbor spin-pair from the noise-source, obtained from exact numerical simulations,  in contrast  to the linear one as predicted by the Lieb-Robinson theorem.    
\end{abstract}

\maketitle

\section{Introduction}
\label{sec:intro}

Rapid development of quantum information technology  has been possible due to the path-breaking inventions of communication  and computational  schemes, including  classical information transmission via quantum states with or without security \cite{bennett92,zukowski98,ekert91}, quantum state transfer  \cite{bennett93,sende10},  quantum metrology \cite{pezze09}, and one-way quantum computation \cite{briegel09}.  An almost universal feature in all these quantum information tasks is the use of quantum correlations in the form of entanglement \cite{horodecki09} between the constituents of composite quantum systems as resource. Over last few years, highly entangled bipartite and multipartite states  have been created in the laboratory using different substrates like photons \cite{raimond01}, trapped ions \cite{leibfried03},   superconducting materials \cite{barends14}, nuclear magnetic resonances (NMR) \cite{vandersypen05}, and optical lattices \cite{mandel03}, making the implementation of quantum information processing protocols using few qubits possible.

A main obstacle in this enterprise is the fragility of entanglement to decoherence \cite{yu09}, which is exhibited by the rapid decay of entanglement with time in multiparty quantum systems exposed to noisy environments \cite{rivas14,petruccione}. This restrains the success of realizing quantum information schemes like transmission of information through quantum channels and implementation of quantum gates with high fidelities. One of the extensively studied scenarios of noisy environments is the consideration of local perturbation in the system due to the Markovian environmental interactions \cite{rivas14,petruccione}. Here, the perturbation lasts for a small time interval, $\delta t$, which is  infinitesimally small compared to our observational time scale, and as per the Markovian approximation, at the beginning of the next time interval, the state is again set to be a product state between the system and the environment, so that the memory effect in the system is not taken into consideration. It has been shown,  both theoretically and experimentally, that entanglement in a multiparty system decays fast, and can even completely disappear after a finite period of time, when subjected to such local environments \cite{yu09}. In contrast, under carefully specified initial conditions, quantum correlations \cite{modi12} such as quantum discord \cite{henderson01}, which are independent of entanglement, may exhibit robustness against similar environmental effects \cite{werlang09}, and can even be preserved for some time \cite{mazzola10}. However, despite a few attempts \cite{carnio15}, realizable situations for preserving entanglement,  as yet, remains elusive.

With this motivation, we present scenarios involving realizable physical systems and environmental models in which entanglement of the system, even when exposed to the environment, remains constant for a finite interval of time at the beginning of the dynamics. We call this phenomena as \emph{freezing} of entanglement.  In recent times, a wide spectrum of substrates is probed in the laboratories all over the world, thereby providing a large set of physical systems to search for the frozen entanglement. Apriori, it is not at all clear which of these systems are more preferable for exhibiting such phenomena in comparison to the others. In this respect, we find that
low-dimensional quantum spin models (QSMs), which can be realized and controlled in different physical systems, including ion traps \cite{porras04}, optical lattices\cite{duan03}, solid-state materials \cite{schechter08}, NMR \cite{zhang12}, and superconducting qubits \cite{dalmonte15}, 
 stand out as excellent candidates. 

In this paper, we consider a local dissipative Markovian noise model in the form of a local repetitive quantum interaction (LRQI) \cite{attal06,dhahri08} (cf. \cite{karevski09}). Such a scenario can be observed in  two physical situations. One of them is repeated applications of quantum measurements \cite{attal06,deb_12}, where identical measurement devices are operated repeatedly, one after another, on the system or parts of the system, while the second one can be seen in quantum optical devices, where a sequence of independent atoms arrives and interacts, one atom after the other, with a quantized radiation field in a cavity  for a short period of time due to the finite life-time of atoms \cite{englert02,deb_345}. Apart from these two scenarios, LRQI is also relevant in electronic transport \cite{deb_6}, thermalization \cite{deb_7}, etc.  We also consider a non-dissipative noise model, represented by the local dephasing noise \cite{petruccione,luczka90,haikka13}, which can arise due to a fluctuation in the external electromagnetic field \cite{carnio16}.  

More specifically, we consider a number of paradigmatic one-dimensional (1D) QSMs defined on spin-$1/2$ particles as systems, namely, the anisotropic XY model in external uniform as well as alternating  transverse fields \cite{divakaran16,thermal_chanda,dutta15, xy1,xy2} (cf. \cite{dm_16}), 
the XYZ model \cite{xyz1,xyz2,xyz3,xyz4} including
the XXZ model with and without an external magnetic field \cite{bethe31,takahashi99,fm_xxz}, and the $J_1-J_2$ model \cite{majumdar69}. We focus on a situation where the local environments interact with one, or more than one selected spins in the system via local repetitive quantum interaction, or by local dephasing. Such a situation may arise in a quantum computer architecture in which only some parts of the system are exposed to the environment and moreover, those exposed parts are such that they cannot be deleted from the system. The inability of deleting parts of a system can, for example, occur in  nuclear magnetic resonance (NMR) molecules and solid state systems.

We show that bipartite entanglement, as quantified by the logarithmic negativity \cite{defLN,vidal02,peres96} over the nearest-neighbor spin pairs in one-dimendional quantum spin systems, freezes for both the dissipative and the non-dissipative noises. This is observed for all nearest-neighbor spin-pairs in the system except for the spin-pair(s) that is (are) adjacent to the environment(s). Freezing of entanglement exists in all the phases of the model, while the length of the freezing duration, corresponding to a chosen nearest-neighbor spin-pair, depends on the choice of the system parameters. We also show that the duration of freezing corresponding to a specific spin-pair in the spin-chain may remain unaffected by a variation of the system size, thereby exhibiting a \emph{scale invariance}. We test the effect of an increase in the temperature of the environment, and introduction of disorder \cite{disorder,disorder-experiment} in the system, and find that the freezing of entanglement is qualitatively robust against such disturbances. We demonstrate how the freezing  of entanglement disappears when the number of system-spins affected by the external environments are increased. We also discuss the relation between the freezing phenomena with the Lieb-Robinson theorem \cite{LR1} on the propagation of information through quantum many-body systems, and point out that the actual values of freezing-duration are considerably higher than the same predicted by the Lieb-Robinson theorem, thereby indicating a much slower propagation of noise through the system particularly when the system size increases.

The paper is organized as follows. In Sec. \ref{sec:method}, we discuss the quantum spin models, and provide a brief description of the different noise models considered in this paper. Sec. \ref{sec:freezing} contains the results on the freezing phenomena of entanglement, including its scale-invariance (Sec. \ref{sec:scale_inv}),  robustness against thermal noise and disorder in the system (Sec. \ref{sec:robustness}), and its connection to Lieb-Robinson theorem (Sec. \ref{sec:lr}). Sec. \ref{sec:conclude} contains concluding remarks.

\section{Models and Methodology}
\label{sec:method}

In this section, we discuss the important features of the relevant quantum spin models used in this paper. We also provide a brief description of the dissipative local repetitive quantum interaction model and local dephasing noise considered in this paper.   

\subsection{The system}
\label{subsec:syst}

To exhibit the freezing phenomena, we consider
a class of generic 1D QSMs constituted of $L$ spin-$\frac{1}{2}$ spins 
with open boundary conditions (OBC) as system.
It is described by the Hamiltonian, $H_S$, given by
\begin{eqnarray}
H_S &=&\sum_{i=1}^{L-1}\frac{J}{4}\Big[(1+\gamma)\sigma^i_x\sigma^{i+1}_{x}
           +(1-\gamma)\sigma^i_y\sigma^{i+1}_{y}\Big]\nonumber\\
           &&+\sum_{i=1}^{L-1}\frac{J \Delta}{4}\sigma^i_z\sigma^{i+1}_z+\sum_{i=1}^L\frac{1}{2}\Big[h_1+(-1)^ih_2\Big]\sigma^i_z.
\label{eq:hamiltonian}           
\end{eqnarray} 
Here,  $\sigma_\alpha$, $\alpha=x,y,z$, are the Pauli matrices, $J >0$ is the strength of the exchange interaction between nearest-neighbor (NN) spins, while $\gamma$ and $\Delta$ are the $x-y$ and the $z$ anisotropies respectively. The system is in the presence of a transverse uniform magnetic field of strength $h_1$, and a transverse site-dependent magnetic field, having strength $h_2$, that changes its direction from $+z$ to $-z$ depending on whether the lattice site is even, or odd. For $\Delta=0$, $H_S$ describes an 1D alternating-field anisotropic $XY$ model (ATXY)  \cite{divakaran16, thermal_chanda,dutta15}.  Other paradigmatic QSMs emerging out of Eq. (\ref{eq:hamiltonian})  are (i) the 1D transverse-field XY model (TXY) ($h_2/J=0,\Delta=0$), (ii) the fully isotropic 1D Heisenberg model ($\gamma=0,\Delta=1,h_2/J=0$), (iii) the 1D anisotropic XXZ model in an external uniform magnetic field (TXXZ) ($\gamma=0,h_2/J=0$) \cite{bethe31,takahashi99}, and (iii) the 1D XYZ model in a uniform magnetic field (TXYZ) ($\gamma\neq0,h_2/J=0$).

For the purpose of demonstration, we use the ATXY and the TXXZ models. We choose the ATXY model over the widely studied TXY model due to the richer phase diagram of the former, where an antiferromagnetic (AFM) and two paramagnetic (PM-I and PM-II) phases appear \cite{dimertoPMII}. In the thermodynamic limit and with the periodic boundary condition (PBC), the phase boundaries of the ATXY model are given by
\begin{eqnarray}
(h_1/J)^2 &=& (h_2/J)^2+1  \ \ \ \  \ \ (\mbox{PM-I} \leftrightarrow \mbox{AFM}), \nonumber \\
(h_2/J)^2 &=& (h_1/J)^2+\gamma^2 \ \ \ \  (\mbox{PM-II} \leftrightarrow \mbox{AFM}) 
\end{eqnarray} 
on the $(h_1/J,h_2/J)$-plane \cite{divakaran16,dutta15} (cf. \cite{dimertoPMII,dm_16}). For OBC,  we observe that the phase boundaries change only slightly, even with a moderately small system size, and the AFM region shrinks.  

On the other hand, the TXXZ model also shows three phases, namely, an AFM, a ferromagnetic (FM), and an XY (spin 
flopping) phases, among which the first two are gaped, while the third one has a gapless spectrum. Specifically, without the external magnetic field, the FM$\leftrightarrow$XY transition occurs at $\Delta=-1$, while at $\Delta=1$, the XY$\leftrightarrow$AFM transition takes place. With increasing the strength of the external field, the quantum phase transition points, $\Delta_c = \pm 1$, shifts to the left (see \cite{takahashi99} for the phase diagram of the model). Here, we point out that in the FM phase ($\Delta \leq -1$), the bipartite entanglement vanishes for all values of the external field \cite{fm_xxz}.

\begin{figure*}
\includegraphics[width=0.7\textwidth]{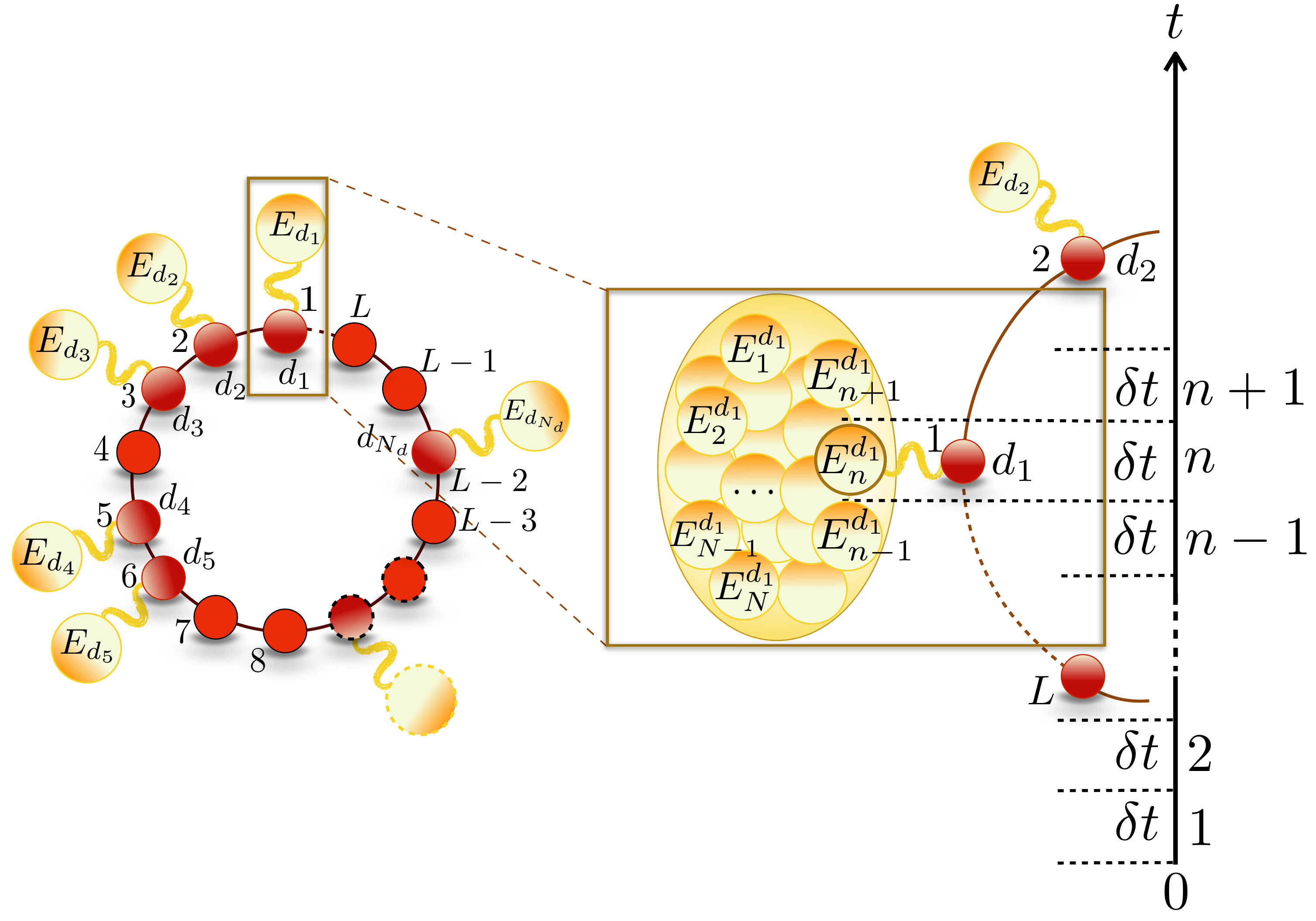}
\caption{(Color online.) Schematic representation of a 1D system of $L$ spins, of which $N_d$ spins, labelled as $d_i$, act as the doors, and interact with independent environments, denoted by $E_{d_i}$.  The enlarged portion describes the local repetitive interaction between the environment and a door  in the system. The spin ``$d_1$'' in the 1D QSM acts as the door, and interacts with a copy of the environment for a short interval of time $\delta t$. In the $n^{th}$ interval of duration $\delta t$, the interacting copy of the environment is $E_n^{d_1}$. Note here that during the same $n^{th}$ interval of duration $\delta t$, along with the door $d_1$, the door $d_i$ in the system $(i\neq 1)$ is also interacting with the copy $E_{n}^{d_i}$ of the environment.}
\label{fig:schematics}
\end{figure*}

\subsection{The environments}
\label{subsec:env}

Let us now consider the situation where at time $t=0$,  $N_d$ number of spins, labeled as $\{d_1,d_2,\cdots,d_{N_d}\}$ (see Fig. \ref{fig:schematics}), from the system, $S$, start interacting with local environments, denoted by $E_{d_i}$. We call these spins in the system to be the ``doors", and consider the type of  interaction between the door and the environment to be Markovian.  The time-evolution of the state of the system, $\rho_S(t)$, is then given by the solution of the Lindblad quantum master equation \cite{rivas14,petruccione}
\begin{eqnarray}
\frac{d\rho_S}{d t} = - \frac{i}{\hbar} [H_S, \rho_S] +\mathcal{D}(\rho_S).
\label{eq:qme}
\end{eqnarray} 
We assume that the environments $\{E_{d_i}\equiv E\}$ are identical, and are independent of each other. The dynamical term $\mathcal{D}(.)$ in Eq. (\ref{eq:qme}) depends explicitly on the physical nature of the environment(s) and the type of the interaction(s) between the door(s) and the environment(s). We now briefly describe the different noise models, corresponding to the different types of environments considered in this paper.

\subsubsection{Local repetitive quantum interaction}
\label{subsubsec:lrqi}

We first consider a dissipative noise model, and start with the scenario in which where there is only one door spin, denote by $d$, in the system. Consider the system, $S$, characterized by the canonical equilibrium state $\rho_S$, to be at absolute temperature $T_S$.  The system, via the door, is in contact with a bath in the form of a collection of $N$ identical and decoupled spins, denoted by $\{E_1^d,E_2^d,\ldots,E_N^d\}$, where $N$ is a large number. To keep the notations uncluttered, we shall discard the superscript ``$d$" in the case of the single door scenario, and denote the spins in the bath as $\{E_1,E_2,\ldots,E_N\}$. However, in the multiple bath scenario to be considered in subsequent discussions, the bath spins corresponding to the door spin $d_i$ are denoted by $\{E_1^{d_i},E_2^{d_i},\ldots,E_N^{d_i}\}$ (See Fig. \ref{fig:schematics}). Each spin in the collection is at absolute temperature $T_E$, and is described by the Hamiltonian $H_{E_i}=B\sigma^z_i$ in the Hilbert space $\mathcal{H}_{E_i}$.  We consider the system-environment ($SE$) interaction to be such that $S$ interacts with only one chosen spin, say, $E_i$, at a given time instant, and the interaction lasts for a very short time-interval, $\delta t$. During this interval, all the other spins in the collection, $\{E_j, ~ j\neq i\}$, remain isolated from $S$ as well as from $E_i$. The total Hamiltonian, $H_i$, describing altogether the combination of the system, $S$, the spin from the collection, $E_i$, with which $S$ interacts,  and the interaction between $S$ and $E_i$,  is defined in the Hilbert space $\mathcal{H}_S\otimes\mathcal{H}_{E_i}$.

Without any loss of generality, we assume that during the first interval $[0,\delta t]$, $S$ interacts with $E_1$. The duo of $S$ and $E_1$, denoted by $SE_1$, has the state $\rho^0_{SE_1}=\rho_S^0\otimes\rho_{E_1}$ at $t=0$, where $\rho_S^0$ is the state of $S$ at $t=0$, and $\rho_{E_1}$ is the state of the spin $E_1$ at temperature $T_{E}$. The unitary evolution generated by $H_1$ in the interval $[0,\delta t]$ is given by $\rho_{SE_1}^0 \mapsto \rho_{SE_1}^1=\mathbb{U}_1 \rho_{SE_1}^0 \mathbb{U}_1^{\dagger}$, where $\mathbb{U}_1 = \exp(-i \delta t H_1/\hbar)$. In the next interval $[\delta t,2\delta t]$, the system, having an initial state $\rho_S^1=\text{tr}_{E_1}\left[\rho_{SE_1}^1\right]$, interacts with $E_2$ only,  and the initial state of $SE_2$ is given by $\rho_S^1\otimes\rho_{E_2}$. In this interval, the dynamics is governed by the Hamiltonian $H_2$, which is defined in a way similar to $H_1$. Note here that $\rho_{E_1}$ and $\rho_{E_2}$ are identical to each other. Continuing this procedure in all subsequent intervals is equivalent to a local repetitive interaction between the system $S$ and one spin, denoted by $E$ and defined by the Hamiltonian $H_E=B\sigma^z_E$, which interacts with the system via the door. At the beginning of every time interval, the initial state of the system-environment duo, $SE$, is reset to the product of the state of the environment, $\rho_E$ (which is the Markovian approximation) and the evolved state of $S$, obtained by tracing out the environment from the evolved state of $SE$ at the end of the previous interval.

In this paper, we consider the interaction Hamiltonian to be of the form 
\begin{eqnarray}
H_{int}(\delta t) = \sqrt{k/\delta t} \left(\sigma_d^x \otimes \sigma_E^x + \sigma_d^y \otimes \sigma_E^y\right),
\label{eq:h_int}
\end{eqnarray}
where the subscript ``$d$" denotes the single door in the system, and $k$ has the dimension of (energy${^2} \times$ time). The total Hamiltonian of the system and the environment is of the form 
\begin{eqnarray}
H = H_S \otimes \mathbb{I}_E + \mathbb{I}_S \otimes H_E + H_{int}(\delta t). 
\label{eq:h_tot}
\end{eqnarray} 
In a single door scenario, this leads to a dynamical term of the form (see Appendix \ref{app:lrqi} for a detailed derivation)
\begin{eqnarray}
\mathcal{D}_d(\rho_S)&=& \frac{2 k}{\hbar^2}\sum_{l=0}^1 p_l [2\eta_{d}^{l}\rho_S\eta_{d}^{l+1}-\{\eta_{d}^{l+1}\eta_{d}^{l},\rho_S\}], 
\label{eq:dynamical_term}
\end{eqnarray}
with $p_0=Z_E^{-1}\exp\left({- \beta_E B}\right)$, $p_1 = Z_E^{-1}\exp\left({\beta_E B}\right)$,  $Z_E=\mbox{tr}[\exp (-\beta_E H_E)]$, and $\eta_{d_i}^\alpha= \left(\sigma_{d_i}^x+i(-1)^\alpha\sigma_{d_i}^y \right)/2$. The operator $\mathcal{D}_d(.)$ reduces to that corresponding to the well-known \emph{amplitude-damping} noise \cite{petruccione} in the limit of high $B\beta_E$.

\subsubsection{Local dephasing noise}

The second type of noise that we consider is the non-dissipative local dephasing noise on $N_d$ of the parties in $S$, thereby leading to a collective dephasing of the chosen parties. Each door, $d$, experiences a pure dephasing noise, being in contact with a thermal bath of harmonic oscillators with frequencies $\{\omega_i\}$,  defined by the Hamiltonian $H_E=\sum_i\omega_ia_i^\dagger a_i$. Here, $a_i(a_i^\dagger)$ is the annihilation (creation) operator of the $i^{\mbox{\small th\normalsize}}$ mode. The interaction Hamiltonian is given by $H_{int}=\sum_i\sigma_d^z\otimes(g_ia_i+g_i^* a_i^\dagger)$, $g$ being the door-reservoir coupling constant. Assuming the zero-temperature state to be the initial state of the reservoir \cite{luczka90},  in a single-door scenario, the dynamical term is given by 
\begin{eqnarray}
\mathcal{D}_d(\rho_S)=\tilde{\gamma}(t)\big(\sigma_d^z\rho_S\sigma_d^z - \rho_S\big),
\label{eq:dyn_term_dephase}
\end{eqnarray}
with
\begin{eqnarray}
\tilde{\gamma}(t)= \omega_c [1  + (\omega_c t)^2]^{-s/2} \sin(s \tan^{-1}(\omega_c t)) \int_0^{\infty} x^{s-1}e^{-x} dx\nonumber\\
\end{eqnarray}
being the zero-temperature time dependent dephasing rate. Here, $\omega_c$ is the cut-off spectral frequency,   and $s$ is the Ohmicity parameter \cite{haikka13}, determining Markovianity $(s\leq2)$.

At this point, it is logical to look into the effect of the presence of multiple doors in the system, and the situation where more than one independent environments are interacting with the same door spin in the system.  The fact that the environments interacting with different doors in the system are independent of each other implies their effect to be additive, which leads to the dynamical term of the multiple-door system with $N_d$ doors given by 
\begin{eqnarray}
\mathcal{D}(\rho_s)=\sum_{i=1}^{N_d}\mathcal{D}_{d_i}(\rho_s),
\label{eq:mult_door}
\end{eqnarray}
where $\mathcal{D}_{d_i}(\rho_S)$ are of the form given in Eq. (\ref{eq:dynamical_term}) or Eq. (\ref{eq:dyn_term_dephase}), depending whether the noise is of LRQI or the dephasing type. One may also consider a scenario where not one, but a finite number, $r_{d_i}$, of environments interact independently on the door $d_i$ during each time interval $\delta t$. Again, these environments being independent of each other lead to a simple modification of Eq. (\ref{eq:mult_door}) as
\begin{eqnarray}
\mathcal{D}(\rho_s)=\sum_{i=1}^{N_d}r_{d_i}\mathcal{D}_{d_i}(\rho_s).
\label{eq:sup_mult_door_2}
\end{eqnarray}

\begin{figure*}
\includegraphics[width=\textwidth]{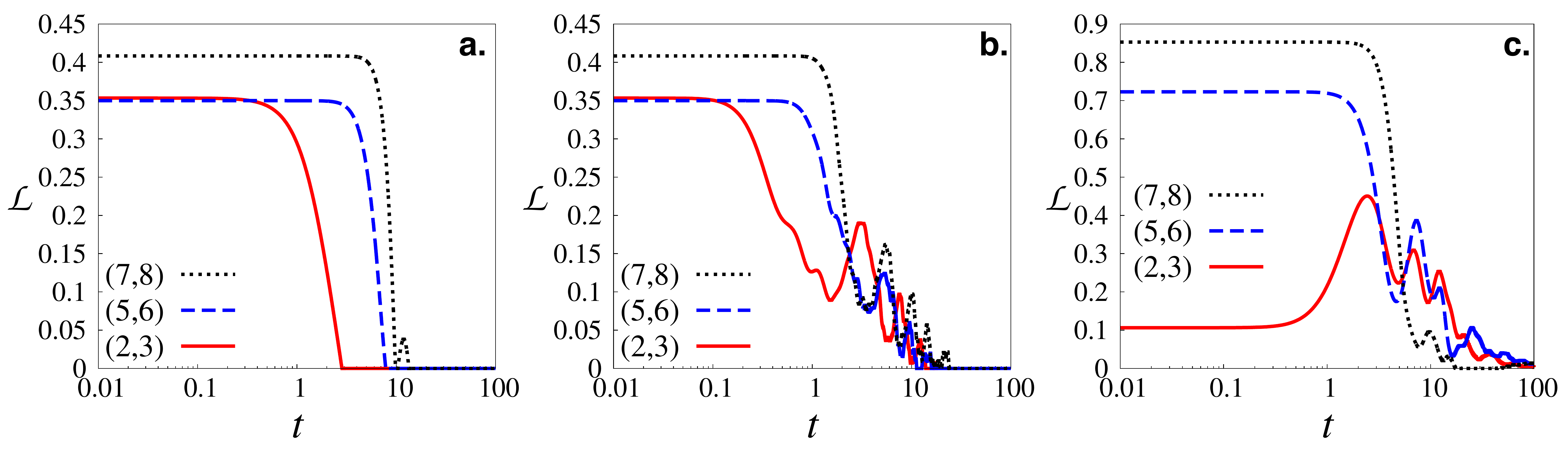}
\caption{(Color online.) \textit{Freezing dynamics of NN entanglement.} \textbf{a.} The NN entanglement freezes in the PM-II phase of the ATXY model for dissipative local repetitive quantum interaction and \textbf{b.} for local phase-damping noise, where the time-axis is in log scale, and the system parameters used in this figure are given in Table \ref{tab:values}.  \textbf{c.} Similar dynamics is observed in the case of the AFM phase in the TXXZ model, where we choose $\Delta=1.5$ and $h_1/J=0.1$. All the axes in all the figures are dimensionless.}
\label{fig:dyn}
\end{figure*}

\section{Freezing of entanglement}
\label{sec:freezing}

In this section, we discuss the main result of this paper, namely, the freezing of NN bipartite entanglement, as measured by \emph{logarithmic negativity} (LN) \cite{defLN, vidal02} in quantum spin models.
Note that the results obtained here remain qualitatively unaltered if one considers other bipartite entanglement measures like entanglement of formation \cite{eof_ref}, concurrence \cite{conc_ref} etc.
 We evaluate LN  of the time-evolved state, $\rho_{i,i+1}(t)=\mbox{tr}_{i,i+1}(\rho_S(t))$, of any two NN spins $(i,i+1)$, $i=1,2,\ldots,L-1$, denoted by $\mathcal{L}_{i,i+1}(t)$. Here $\rho_S(t)$ is obtained by solving Eq. (\ref{eq:qme}) via employing the fourth order Runge-Kutta method, for which the order of the local numerical errors goes as the fifth-power of the length of increment of time in each iteration step of the algorithm. For our purpose, we set the length of increment in time as $0.01$, such that the local numerical error is $\sim 10^{-10}$. We consider a canonical equilibrium state 
\begin{eqnarray}
\rho_S^0=\frac{\exp(-\beta_S H_S)}{\mbox{Tr}[\exp(-\beta_S H_S)]} 
\end{eqnarray}
of $S$ at absolute temperature $T_S$ as the initial state. Let us denote the value of $\mathcal{L}_{i,i+1}$ at $t=0$ by $\mathcal{L}^0_{i,i+1}$. We consider $\mathcal{L}_{i,i+1}(t)$ to be frozen over a time interval $[0,\tau_F^{i,i+1}]$, $0\leq\tau_F^{i,i+1}\leq t_l$, if for all  $t$ in $[0,\tau_F^{i,i+1}]$, 
\begin{eqnarray}
\left|\mathcal{L}_{i,i+1}(t)-\mathcal{L}^0_{i,i+1}\right|\leq\delta;\,\mathcal{L}^0_{i,i+1}>0,
\label{eq:freezing}
\end{eqnarray} 
where we choose $\delta$ to be $10^{-5}$. We call $\tau_F$ to be the \emph{freezing terminal}, which is a characteristic of the chosen NN spin-pair as well as  the  parameters defining the system, the environment, and the system-environment interaction. The typical value of the quantity $t_l$ is large, and has to be chosen by a careful inspection of LN.  A time-span, $t_l$, is considered to be large if LN saturates to a fixed value for $t\geq t_l$, due to the equilibration of the system, or, for instance, some accidental cancellations within the expressions representing LN, which is not necessarily equivalent to the equilibration of the entire system. In the present case, $t_l \sim 10^3$. A dimensional analysis of Eq. (\ref{eq:qme}), taking into account the form of the system Hamiltonian given in Eq. (\ref{eq:hamiltonian}), leads to defining the dimensionless quantities, $k \rightarrow k/(\hbar J)$, $t \rightarrow Jt/\hbar$, $\beta_S\rightarrow J\beta_S=J(k_B T_S)^{-1}$, and $\beta_E\rightarrow B \beta_E=B(k_B T_E)^{-1}$,  used throughout this paper, where we set $k = 1$ for all our calculations. 

\begin{table}
\begin{tabular}{ |c|c|c|c|c| } 
 \hline
 Phase & Specimen values & $\tau_F^{i,i+1}$ vs. $i$ & SI \\ 
 \hline
PM-I & $\frac{h_1}{J}=1.2,\,\frac{h_2}{J}=0,\,\gamma=0.8$ & M & All \\ 
 \hline 
 PM-II & $\frac{h_1}{J}=0,\,\frac{h_2}{J}=1.2,\,\gamma=0.8$ & M & All \\
 \hline
 AFM & $\frac{h_1}{J}=0.2,\,\frac{h_2}{J}=0.2,\,\gamma=0.8$ & NM & Selective \\ 
 \hline
\end{tabular}
 \caption{Values of the system parameters chosen for demonstration in different phases of the ATXY model. The last two columns indicate the type of variation (monotonic (M) or non-monotonic (NM)) of $\tau_F^{i,i+1}$ with $i$, and whether all the spin-pairs show scale-invariant (SI) freezing in the phase (See Fig. \ref{fig:scale} and discussions in Sec. \ref{sec:scale_inv}). Note, however, that the results reported here is true even for other system- and environment-parameters. All parameters are dimensionless.}
 \label{tab:values}
\end{table}

\begin{figure}
\includegraphics[width=0.5\textwidth]{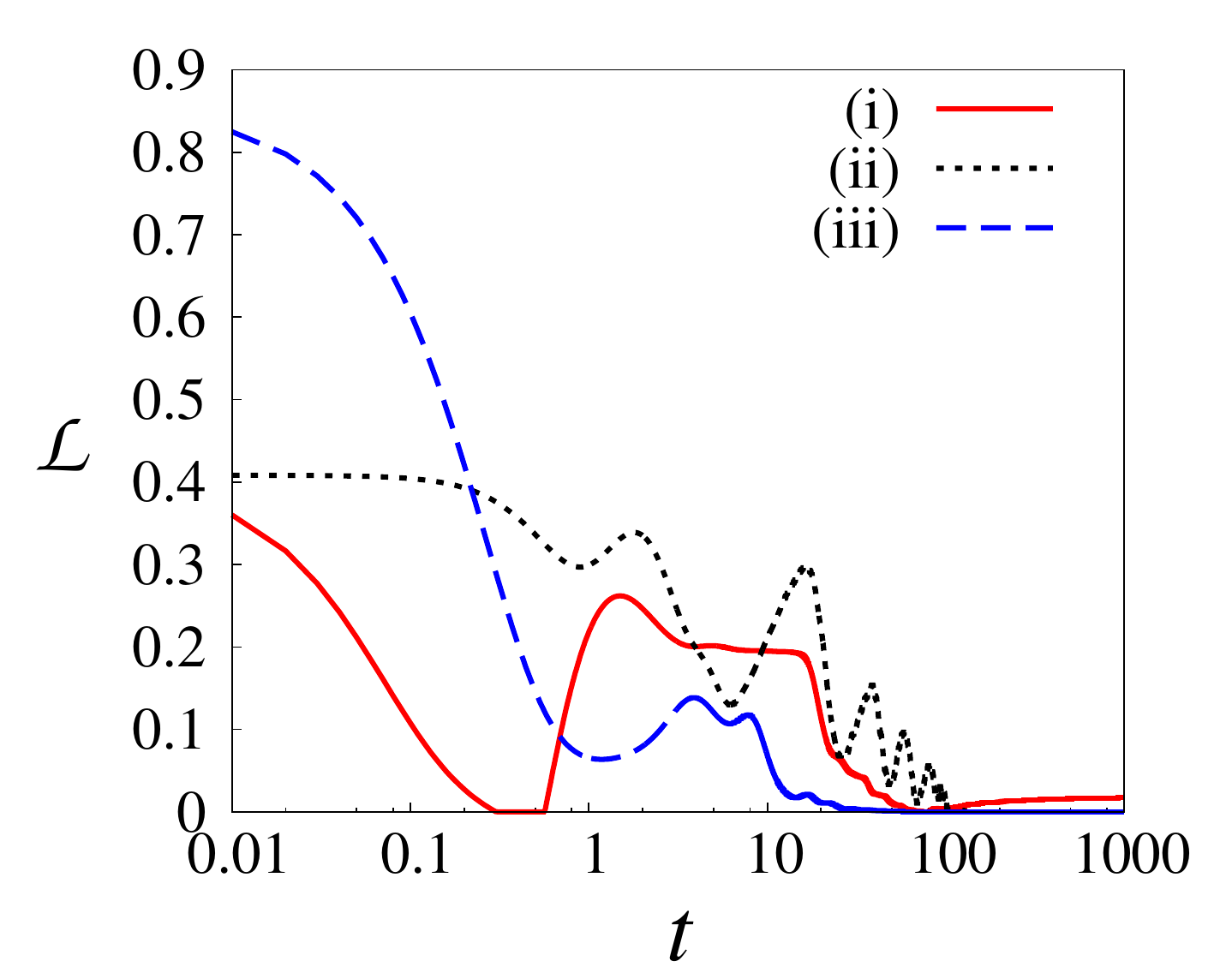}
\caption{(Color online.) Time-dynamics of entanglement of the spin-pair $(1, 2)$  for (i) PM-II phase of the ATXY model under LRQI and (ii) under local dephasing, where model-parameters are given in Table \ref{tab:values}, as well as for (iii) AFM phase of the TXXZ model under LRQI, where we choose $\Delta = 1.5$ and $h_1/J = 0.1$.
All the axes in the figure are dimensionless.}
\label{fig:1_2}
\end{figure}

For demonstration, we use the LRQI model, and fix $ J\beta_S=20$, and $B \beta_E=10$ for all our calculations. Note here that the value of $B\beta_E>5$ ensures that the LRQI model effectively represents the local Markovian amplitude-damping noise, and our calculations, therefore, are performed in the amplitude-damping regime of the noise model.   In the single-door scenario, we consider the spin ``$1$" as the only door in the system.  The different values of the system parameters used for demonstration, corresponding to different phases of the ATXY model, are tabulated in Table \ref{tab:values}. In all three  phases of the ATXY model, NN entanglement, corresponding to all the spin-pairs except those with a door, remains constant for a finite interval of time. The preservation of entanglement, corresponding to the NN spin-pair $(i,i+1)$, $2\leq i\leq L-1$, occurs at the beginning of the dynamics, thereby exhibiting a freezing  of entanglement with a finite $\tau_F^{i,i+1}$. For $t>\tau_F^{i,i+1}$, $\mathcal{L}_{i,i+1}(t)$, $2\leq i\leq L-1$, decays rapidly to zero with increasing time, and eventually undergoes a sudden death. See Fig. \ref{fig:dyn}{\textbf{a.}} for a demonstration with $L=8$ and $N_d=1$, where spin ``$1$" is chosen as the door in all the figures in Fig. \ref{fig:dyn}. 

\begin{figure*}
\includegraphics[width=\textwidth]{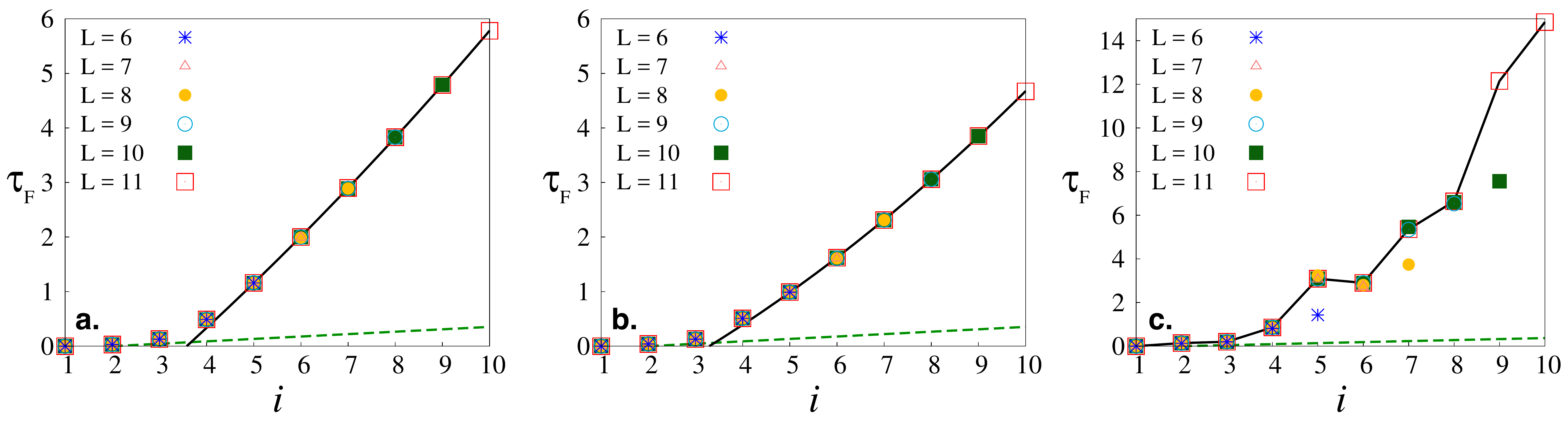}
\caption{(Color online.) \emph{Scale-invariance.} The behavior of $\tau_F^{i,i+1}$ against $i$, for $6\leq L\leq 11$, in the \textbf{a.} PM-I, \textbf{b.} PM-II,  and \textbf{c.} AFM phases of the ATXY model, with the chosen system parameters given in Table \ref{tab:values}. Different point-types correspond to different values of $L$. In \textbf{a.-b.}, the points corresponding to $L=11$ are joined by a continuous line, which clearly exhibits the monotonicity, while such monotonic behavior is not present in case of \textbf{c.}.  The (green) dashed curves, in all the figures, show the variation of freezing terminal, $\tau_F^{LR}$, as predicted by the Lieb-Robinson theorem, with $i$ (see discussions in Sec. \ref{sec:lr}). All quantities plotted are dimensionless.}
\label{fig:scale}
\end{figure*}

Note here that the \emph{freezing} phenomena is different than \emph{saturation} \cite{carnio15} (cf. \cite{apollaro10}) since the latter occurs only at large time, while the former takes place right after the system starts interacting with the environment. It is important to stress here that the other noisy environments, inevitably present in experiments, and   usually ignored in theoretical studies, will increase their effects on entanglement of the system at later times, which may disturb the saturation phenomena  while such possibilities are reduced in  freezing of entanglement. Note also that in contrast to the Markovian system-environment interaction, there exists instances of revival of LN after a complete collapse to zero (Fig. \ref{fig:dyn}\textbf{a.}). This is a result of the non-zero interaction between the spins in the system at all time during the dynamics, including at $t=0$, which generates a memory effect in the bulk of the system. 

See Fig. \ref{fig:dyn}{\textbf{b.}} for a demonstration of the freezing phenomena under the Markovian dephasing noise with $L=8$ and $N_d=1$. Note here that irrespective of the type of noise, the temperature, $J\beta_S$, of the system at $t=0$ has to be such that $\mathcal{L}_{i,i+1}^0>0$ to satisfy Eq. (\ref{eq:freezing}). 
In this context, it is worthwhile to mention that in  one-dimensional quantum spin models with short-ranged interactions,  pairwise entanglement dies out rapidly as the distance between the spins forming the spin-pair under consideration increases. In the case of the ATXY model, entanglement for the spin-pairs $(i, i+m)$ with $m>1$ for the thermal as well as the ground state is non-zero at $t = 0$ only for some specific parameter ranges. We find that if entanglement is present in the spin-pair $(i, i+m)$ with $m > 1$, then freezing of entanglement takes place if $i > 1$. In case of PM-II phase of the ATXY model with open-boundary condition (c.f. \cite{osborne} for periodic-boundary condition) with system-parameter values given in Table \ref{tab:values}, entanglement is non-zero only for the spin-pairs $(1,3)$ and $(L-2, L)$ (i.e., when $m=2$) apart from the cases of $m=1$  (nearest-neighbor pairs). Similar to the nearest-neighbor pair $(1,2)$, entanglement for the spin-pair $(1,3)$ does not freeze, while for the pair $(L-2, L)$, freezing of entanglement takes place. Interestingly, we find that the value of freezing terminal ($\tau_F$) for the pair $(L-2, L)$ is larger than the same for the pair $(L-2, L-1)$, but smaller than  that of the pair $(L-1, L)$.

Keeping the model for system-environment interaction unchanged at either the LRQI or the dephasing noise, we observe that the freezing of NN entanglement occurs in the AFM and PM phases of  the TXY model,  in the AFM and the XY phases of the TXXZ model \cite{takahashi99} (see Fig. \ref{fig:dyn}\textbf{c.}), and in TXYZ, fully isotropic Heisenberg, and the 1D $J_1-J_2$ models \cite{majumdar69}. The last model is represented by the Hamiltonian, 
 having an additional next-nearest neighbor interaction term,
\begin{eqnarray}
H_S=J_1\sum_{i=1}^L\vec{\sigma}_i.\vec{\sigma}_{i+1}+J_2\sum_{i=1}^L\vec{\sigma}_i.\vec{\sigma}_{i+2},
\end{eqnarray}
$J_j (j=1,2)$ are coupling constants of nearest neighbor and next-nearest neighbor interactions.
Note that in the TXXZ model with OBC, the freezing phenomena is present in all the phases of the model as depicted in Fig. \ref{fig:dyn}\textbf{c.}, where the system parameters are chosen from the AFM phase of the TXXZ model $(\Delta = 1.5, h_1/J = 0.1)$,
except the FM phase, where bipartite entanglement vanishes at $t = 0$ due to the alignment of the spins, and and remains so when the system interacts with the environment, thereby violating Eq. (\ref{eq:freezing}).These findings emphasize the potential of the freezing phenomena to be generic to the phases of the 1D QSMs. However, in the rest of the paper, we shall focus on the ATXY model to demonstrate the different features of the freezing of entanglement. 
  
 \noindent\textbf{Note.}  As mentioned before, freezing of entanglement is observed for all the nearest-neighbor spin pairs in the system, except the spin-pair $(1,2)$.  In fact, $\mathcal{L}_{1,2}(t)$ exhibits a fluctuating behavior (see Fig. \ref{fig:1_2}). Interestingly, depending on the choice of the noise model  and the quantum phases of the spin model  $\mathcal{L}_{1,2}(t)$ either saturates to a finite value (e.g., PM-II phase of the ATXY model under LRQI, where $\mathcal{L}_{1,2}(t\rightarrow t_l)\approx 0.017$), or goes to zero at large time.

\subsection{Scale invariance}
\label{sec:scale_inv}

In both the PM phases of the ATXY model, the value of $\tau_F^{i,i+1}$, for a specific choice of $(i,i+1)$, remains unaffected with a change in the system-size, indicating a \emph{scale-invariance}. 
Specifically, for fixed $(i,i+1)$, 
\begin{eqnarray}
\tau_F^{i,i+1}=t_c \;\;\forall L, 
\end{eqnarray}
where $0\leq t_c\leq t_l$, $t_l\sim10^3$. The equality is up to our numerical accuracy ($\sim 10^{-5}$). As a result, the variations of $\tau_F^{i,i+1}$ against $i$, corresponding to different values of $L$, coincide (Fig. \ref{fig:scale}\textbf{a.-b.}), indicating an invariance of the variation of $\tau_F^{i,i+1}$ with $i$, against varying $L$.  For $i\geq 5$, where the values of $\tau_F$ are considerably high, this variation is a parabolic one, given by 
\begin{eqnarray}
 \tau_F^{i,i+1}= ai^2+bi+c \;\;\forall L,
 \label{eq:scale-invariance}
\end{eqnarray}
irrespective of the value of $L$, where $a,b,$ and $c$ are determined by the system parameters. For instance, in the example shown 
in Fig. \ref{fig:scale}\textbf{a}., $a=1.77\times10^{-2}\pm1.2\times10^{-3}$, $b=6.6\times10^{-1}\pm1.8\times10^{-2}$, and $c=-2.59\pm6.5\times10^{-2}$,  and in case of Fig. \ref{fig:scale}\textbf{b}., $a=2.41\times10^{-2}\pm1.5\times10^{-3}$, $b=3.767\times10^{-1}\pm2.26\times10^{-2}$, and $c=-1.50\pm8.2\times10^{-2}$.  This equation allows one to estimate $\tau_F$ corresponding to $\mathcal{L}_{i,i+1}$ with increasing distance from the door. The importance of the above result lies in the fact that if execution of a quantum information protocol requires certain time period, say, $\tau^\prime_{F}$,  Eq. (\ref{eq:scale-invariance}) provides the estimate of the minimum size of the system, given by $L_m=i_m+1$,  required to attain this value, where $i_m$ is obtained as a solution of Eq. (\ref{eq:scale-invariance}), by using $\tau_F^{i,i+1}=\tau^\prime_F$. Also, in both of the PM-I and the PM-II phases, the freezing terminal, $\tau_F^{i, i+1}$ shows a monotonic behavior with $i$ given by
\begin{eqnarray}
\tau_F^{i,i+1}\geq\tau_F^{j,j+1}\, \forall\,\, 1<j<i\leq L-1,
\end{eqnarray}
and thereby imposing a hierarchy among the different NN pairs in
$\tau_F$.

However, in the AFM phase, scale-invariance is observed for selected NN spin pairs only (Fig. \ref{fig:scale}\textbf{c.}). Therefore, this feature distinguishes between the paramagnetic and the AFM phases of the ATXY model. Moreover, the variation of $\tau_F^{i,i+1}$ with $i$ is  \emph{non-monotonic} in the AFM phase. The existence of the scale-invariance is, however, independent of whether the trend of $\tau_F^{i,i+1}$ with $i$ is monotonic, or non-monotonic (Fig. \ref{fig:scale}\textbf{c.}).  E.g. $\tau_F^{5,6}>\tau_F^{6,7}$, while $\tau_F^{5,6}$ as well as $\tau_F^{6,7}$ are independent of $L$. These observations indicate that the freezing of entanglement can not simply be explained by the attenuation of the decohering power of the environment as one moves away from the door. It also requires an understanding of how the disturbance due to the bath propagates through the quantum spin-chain. We will again address this question at the end of this section.


The entire analysis in this paper is based on the system Hamiltonian  with OBC. The use of PBC, instead of the OBC, imposes a reflection symmetry in the values of $\tau_F$ with respect to $i=\frac{L}{2}$ $\left(\mbox{or} \ \frac{L-1}{2}\right)$, depending on whether $L$ is even (or odd). Hence, it decreases the maximum achievable value of $\tau_F$ compared to the same in a system with OBC. It is also important to point out here that the  scale-invariance of  $\tau_F$ is found in selected NN spin-pairs of all the other 1D QSMs considered in this paper, but the monotonic increase of $\tau_F^{i,i+1}$ with $i$ is also absent
in those spin models.

\begin{figure}
\includegraphics[scale=0.5]{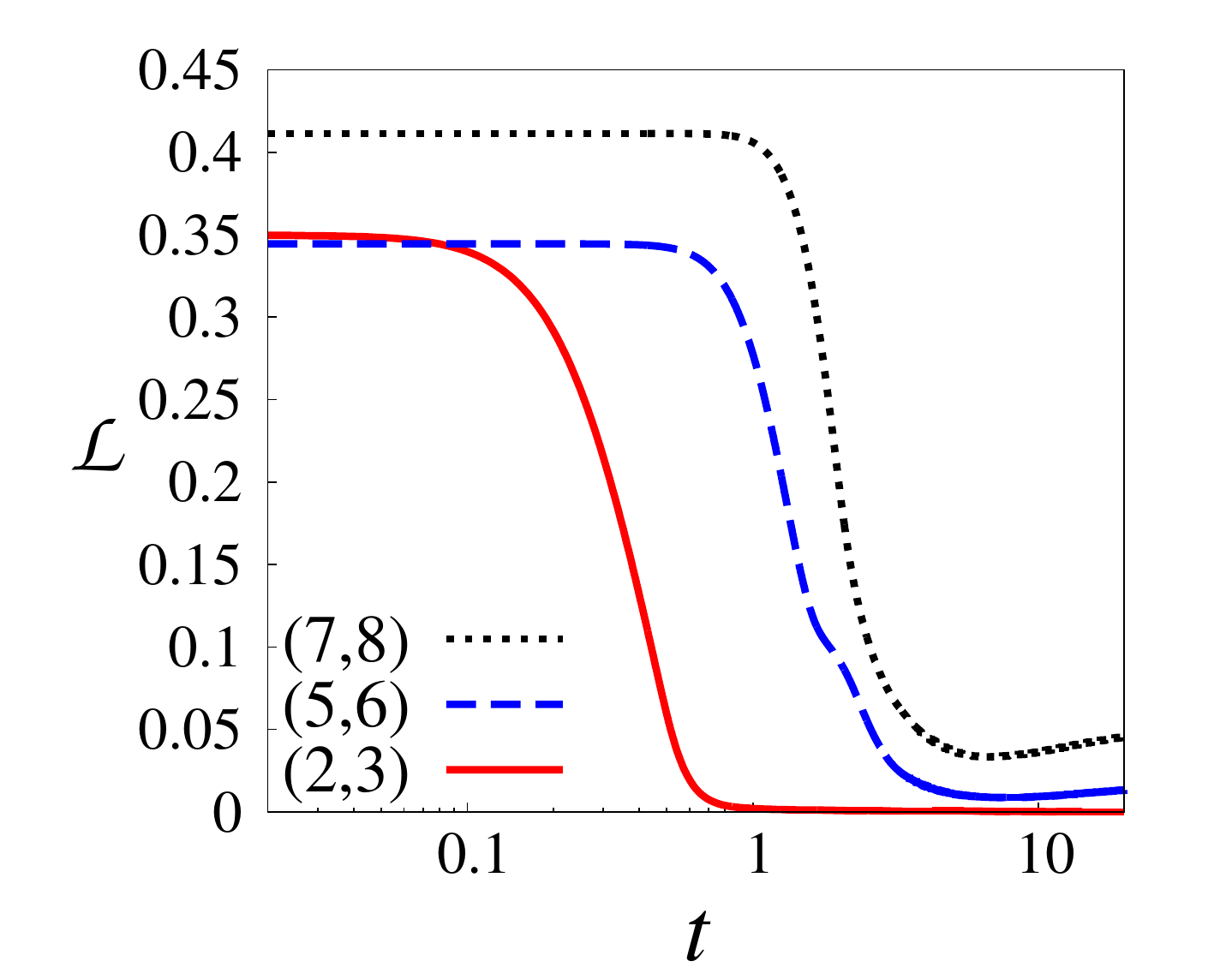}
\caption{(Color online.) Variation of NN quenched entanglement with respect to $t$ under local repetitive interaction in the case of disordered ATXY model, with $h_2^i/J$ as the disordered system parameter. The  values of $h_2^i/J$ are chosen from a Gaussian distribution of mean \(\langle h_2/J \rangle=1.2\) with standard deviation $0.3$, and we set $\gamma=0.8$ and \(h_1^i/J=0\). The time axis is in logarithmic scale, and all the axes are dimensionless.}
\label{fig:dis_dyn}
\end{figure}

\subsection{Robustness}
\label{sec:robustness}

In order to investigate the robustness of the freezing phenomena,
 we consider two specific situations where the system-environment duo with a frozen NN entanglement is subjected to disturbance.  The first situation is that of changing the temperature of the environment from a temperature at which freezing has occurred  in the ATXY model. We find that the qualitative results regarding the freezing of bipartite entanglement, and its scale-invariance, remain unchanged with a change in the environment-temperature, $B\beta_E$, although the entanglement decays more rapidly for $t>\tau_F$ when $B\beta_E$ is low, i.e., when one moves away from the amplitude-damping limit. Similar findings are obtained when one uses a non-dissipative noise, such as the local dephasing noise, instead of a dissipative one.

Next, we also consider a disordered ATXY model, where the strengths of the transverse uniform and alternating magnetic fields, $h_1^i/J$ and $h_2^i/J$, corresponding to the lattice site, $i$, are chosen randomly from Gaussian distributions with  mean  \(\langle h_1/J \rangle\) ,  and \(\langle h_2/J\rangle\), respectively, and with a fixed standard deviation \cite{disorder}, for all the lattice sites. Such systems can now also be engineered in the laboratory with currently available technologies \cite{disorder-experiment}. We assume that the disorder is quenched, where the quenching is performed under the assumption that the time scale of the dynamics is much smaller than the equilibration time of the disorder.  A canonical equilibrium state, corresponding to an initial set of such random values of the system parameter on all the sites at a finite temperature, evolves under the noisy environment. The NN entanglement corresponding to a specific spin-pair at every time instant during the dynamics is computed, and averaged over a large number of initial sets of the values of the chosen system parameter -- we call this average entanglement as the NN quenched entanglement.

Quantum correlations in these disordered systems often show counter-intuitive behavior compared to the corresponding ordered systems \cite{disorder-group-paper}. In the present case, we find that  freezing of NN quenched entanglement occurs with all its qualitative characteristics retained, thereby exhibiting a robustness against  disorder in the system. However, the value of freezing terminal corresponding to a specific spin-pair decreases. An example of the freezing dynamics in the quenched disordered ATXY model is given in Fig. \ref{fig:dis_dyn}, where $h_2^i/J$ is the disordered system parameter, chosen from a Gaussian distribution of mean $\langle h_2/J\rangle=1.2$, and standard deviation $0.3$, with $\gamma=0.8$, and $h_1^i/J=0$  for all lattice sites. Note that in the ordered case, the chosen values of system-parameters are $h_2/J=1.2$, $h_1/J=0$, $\gamma=0.8$ corresponding to the PM-II phase.  The only qualitative difference between the disordered case and the one without disorder is a longer sustenance of entanglement over time, as clearly seen from the figure.

\noindent\textbf{Note.} In Figs. \ref{fig:dyn} and \ref{fig:dis_dyn}, the time axes are in logarithmic scale. Therefore, time has been plotted from $t = 0.01$ instead of $t = 0$. However,  in the interval from $t = 0$ to $t = 0.01$, NN LN remains constant over time  (i.e., frozen) for all the spin pairs $(i, i+1)$, with $i > 1$.

\begin{figure}
\includegraphics[scale=0.7]{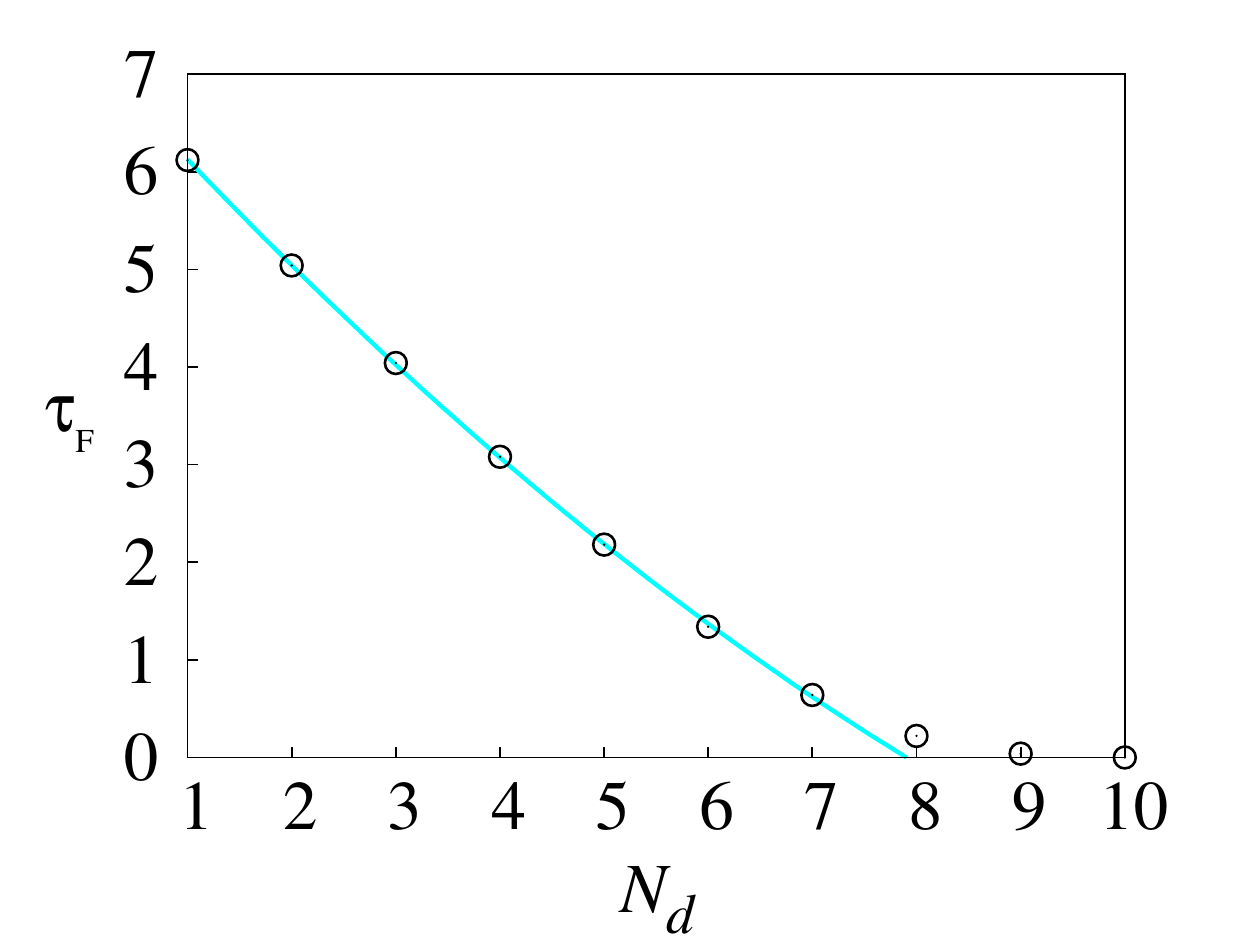}
\caption{(Color online.) Variation of $\tau_F^{10,11}$ with $N_d$ in the PM-I phase of the ATXY model with $L=11$, where doors are added one-by one in the system, starting from spin $1$. All quantities plotted are dimensionless.}
\label{fig:nscale}
\end{figure}

\subsection{Multiple doors and environments}

We now move to the case where instead of one door, the environment affects the system via multiple doors.
We observe that the freezing terminal for a given NN spin-pair in a spin chain of length $L$ decreases when a larger portion of the system is exposed to the environment. For example, in the PM-I phase of the ATXY model, if more doors are added one-by one in the system, starting from spin $1$, $\tau_F^{10,11}$ exhibits a parabolic decay, given by 
\begin{eqnarray}
\tau_F^{10,11}=0.0335714 N_d^2 - 1.18643 N_d + 7.28,
\end{eqnarray}
with increasing number of doors, $N_d$, exposed to the environment (see Fig. \ref{fig:nscale}). Freezing of entanglement entirely vanishes if the entire system is exposed to noise. 

One may also consider a scenario where instead of one, a fixed and finite number of spins, say $r~(>1)$, interact independently as environments with the door $d$ at spin $1$ in the QSM during the same time-interval $\delta t$. 
The effect of each of these $r$ environments is additive (see Sec. \ref{subsec:env} for details). We here find qualitatively similar results regarding freezing of entanglement. However, with increasing $r$, a decrease in the value of the $\tau_F$ is observed. In the AFM phase of the ATXY model, the value of $\tau_F^{i,i+1}$, for a fixed pair of NN spins, decreases monotonically with increasing $r$ approximately as $\sim r^{-1}$. However, in the PM-I and the PM-II phases, and for fixed $(i,i+1)$, both monotonic and non-monotonic variation of $\tau_F^{i,i+1}$  with increasing $r$ are found. The non-monotonic variation of $\tau_F^{i,i+1}$ with $r$ is abundant when one moves away from the phase boundaries.  Also, counter-intuitively, with $r>1$, $\mathcal{L}_{i,i+1}(t)$ for $i>1$ is found to remain non-zero for a longer time after $t>\tau_F^{i,i+1}$, compared to the same in the case of $r=1$, thereby indicating a robustness of entanglement against the increase of the number of environments accessing the system via a single door.

\subsection{Physical interpretation towards freezing of entanglement and Lieb-Robinson velocity}
\label{sec:lr}

One may interprete the freezing terminal, $\tau_F^{i,i+1}$, as the time taken by the disturbance introduced at the door-spin $d$ to reach the spin-pair $(i,i+1)$ situated at a certain distance from the door. Such an interpretation directly connects the freezing phenomena of entanglement with the Lieb-Robinson (LR) theorem \cite{LR1} in many-body physics, which provides upper bounds on the speed of propagation of information in many-body systems.  According to the LR theorem, the speed of information-flow from a subsystem, $X$, to another subsystem, $Y$, of a many-boby system is finite, and is bounded below by the LR velocity $v$ \cite{LR1, LR2}. Therefore, if  $X$ is subjected to a local noise,  its effects will be exponentially suppressed if $d(X, Y) > v t$, where $d(X, Y)$ measures the distance between the subsystems $X$ and $Y$, and $t$ is the time (see Sec. 3 in \cite{LR2} for details). Thus, 
let us
consider the freezing of NN entanglement at a specific NN spin-pair $(i,i+1)$ at a distance $i$ from the door (spin ``1")  to be occurring due to a finite time taken by the  noise at spin ``1" to propagate along the spin-chain to the NN spin-pair. The lower bound of the freezing terminal, according to LR bound, should be $\tau_F^{\text{LR}} \approx d(X,Y)/v$. In our scenario, we apply noise on the spin $1$ which is interacting with spin ``$2$", implying that the subsystem $X$ can be considered as the spin-pair $(1, 2)$. If the freezing of entanglement on the spin-pair $Y = (i, i+1)$ is due to the finite velocity of the effect of noise on $X$ along the spin-chain,  then the freezing terminal $\tau_F^{\mbox{LR}}$,  as estimated from the LR theorem, is given by 
\begin{eqnarray}
\tau_F^{\text{LR}} \approx |i - 2|/v.
\end{eqnarray}
The examples of 1D QSMs used here being of short-range interactions, the variation of $\tau_F^{\text{LR}}$ against $d$ is predicted to be a linear one \cite{LR-initial}, which is indeed the case (see Fig. \ref{fig:scale}).

However, a comparison between the actual value of the freezing terminal and the one obtained by using the LR theorem leads to the following observations. 
\begin{enumerate}
\item Although the LR theorem provides an estimate of the time taken by the noise to travel the distance $d$ through the spin-chain, the actual value is expected to be greater or equal to the LR estimation. Our numerical analysis provides evidence that the actual propagation time of noise is considerably longer than the LR prediction, and is a quadratic function of $d$, which is in contrast to the LR prediction. Although $\tau_F^{\text{LR}}$ and $\tau_F$ may posses values of similar order when $d$ is small, with increasing $d$, the LR estimation of the freezing terminal becomes very small compared to the actual value, thereby predicting a faster propagation of noise, which is actually not the case. Hence in the case of large system size, where the distance between the noise-source and the target spin-pair is large, the LR estimation may become qualitatively different (quadratic vs. linear).  This is clearly demonstrated in Fig. \ref{fig:scale}. 
\item The LR theorem predicts \emph{scale-invariance} of the propagation time in \emph{any} system under consideration, as is clear from expression of $\tau_F^{\text{LR}}$. However, the LR value provides only a lower bound on the freezing terminal. In a specific system, there may exist a scale-invariant freezing terminal at a much higher value than that is provided by LR prediction. But in general, this higher value of freezing terminal is not universally scale-invariant, unlike the LR one,
for example,  in the AFM phase of the ATXY model. 
\end{enumerate}

Our analysis provides an alternative way of investigating the propagation of noise through quantum many-body systems, independent of the LR theorem. It also relates two seemingly different directions of research, namely, the investigation of frozen entanglement under noise and the propagation of information through quantum many-body systems. Moreover,  our analysis clearly demonstrates that the exact analysis may provide results that
have large deviation from
 the LR predictions, and therefore emphasizes the necessity of looking into the actual results even in cases where LR calculations are possible.

Towards understanding the scale-invariance in the freezing phenomena, we study the correlation function
\begin{eqnarray}
C_{ij} =  \langle \vec{\sigma}^i. \vec{\sigma}^j \rangle - \langle\vec{\sigma}^i\rangle \langle\vec{\sigma}^j \rangle,
\end{eqnarray}
where $1 \leq i < j \leq L$, in the QSM at $t=0$. We find that corresponding to the spin pairs $(i,i+1)$ exhibiting scale-invariance (e.g. in the PM-I and PM-II phases of the ATXY model) with spin ``1" as the door, the value of the long-range correlation at $t=0$, given by $C_{1i}$, with $i > 1$, is low compared to the same in the case of spin-pairs that do not exhibit scale-invariance (e.g. selected pairs in the AFM phase of the ATXY model). Moreover, we point out that the correlation length diverges \cite{xy1}
at the phase-boundaries of the ATXY model, where the value of freezing terminal is low. In contrast, well inside the three phases of the model, the value of freezing terminal increases, thereby validating the interpretation of the freezing terminal as the propagation time of disturbance through the spin-chain. Note that while the interpretation seems simple in the case of a single-door system with open boundary condition, for systems with multiple doors and periodic boundary conditions, a chosen spin pair can experience disturbances originating from different doors, thereby indicating an intricate mechanism for the dependence of the freezing duration over the distance of the spin-pair from the door(s).


\section{Conclusion}
\label{sec:conclude}   

Entanglement is known to be an important resource in a large class of quantum information protocols. Therefore, finding robustness of entanglement under different decoherence models has attracted a lot of attention. In this paper, we demonstrated that under local noise, bipartite entanglement of a quantum many-body system can remain constant, or near-constant, within numerical accuracy, over a finite interval of time, called the freezing terminal. 
We call this feature as the freezing of entanglement. 
We  showed that the \emph{freezing} of bipartite entanglement can take place in a collection of paradigmatic one-dimensional quantum spin systems, like
the anisotropic XY model in a transverse uniform and an alternating magnetic field (ATXY),  the XYZ,  and the $J_1-J_2$ models 
under both dissipative and non-dissipative environments.
 As the first kind of noise, we consider a local repetitive quantum interaction, which in the low temperature limit, effectively represents the local amplitude damping noise.  On the other hand, the non-dissipative noise is represented by the local dephasing noise. We showed that freezing of entanglement occurs for both kinds of noise, as well as in all the phases of the quantum spin models considered, except in phases where the bipartite entanglement 
of the initial state
 vanishes, as in the case of the ferromagnetic phase of the TXXZ model. 

We found that in the paramagnetic phases of the ATXY model, the duration of freezing of entanglement, corresponding to all the nearest-neighbor pairs in the system,  is independent of the system-size, thereby exhibiting a scale-invariance. Interestingly, such a scale-invariance was present only in the case of selected nearest-neighbor pairs of spins in the case of the AFM phase of the ATXY model, and in all the phases of the rest of the quantum spin models considered in this paper. We also found that  irrespective of the choice of the quantum spin model, freezing of entanglement remains qualitatively unaffected with a change in the environment-temperature, or in a situation where disorder is introduced  in the system. We also investigated the phenomena where multiple spins in the system was subjected to noise, or when more than one environment interacted with the same spin in the system, and observed the freezing of entanglement to be sustained with qualitative changes only. However, with increasing number of parties in the system that were subjected to noise, the freezing of entanglement eventually vanished.  The quantum spin models as well as the noise models considered in our work can be realized in quantum optical devices, nuclear magnetic resonances and cold atoms in optical lattices, thereby making the realization of frozen entanglement in the laboratory a possible goal. Therefore, our results are expected to have an impact in the making of quantum devices using quantum entanglement as resource.

\acknowledgments 
This research was supported in part by the `INFOSYS scholarship for senior students'. The authors acknowledge computations performed at the cluster computing facility of Harish-Chandra Research Institute, Allahabad, India.

\appendix

\section{Lindblad master equation for local repetitive quantum interaction}
\label{app:lrqi}

Following the description of LRQI in Sec. \ref{subsubsec:lrqi}, let us consider the $n^{th}$ time interval, $[(n-1)\delta t,n\delta t]$, during which the system, $S$, interacts with the $n^{th}$ environment-spin, $E_n$, only, $n=1,2,\ldots,N$. The evolution of the complete state, $\rho$, of the system, $S$, and $N$ copies of the spin, $\{E_i\}$, in this interval is achieved by $\rho\mapsto\tilde{\mathbb{U}}_n\rho\tilde{\mathbb{U}}_n^\dagger$,  where $\tilde{\mathbb{U}}_n$ and $\rho$ are defined in the Hilbert space given by $\mathcal{H}_{tot}=\mathcal{H}_S\bigotimes_{n=1}^N\mathcal{H}_{E_n}$. The operation $\tilde{\mathbb{U}}_n$ is given by 
\begin{eqnarray}
\tilde{\mathbb{U}}_n=\mathbb{U}_n\bigotimes_{\underset{m\neq n}{m=1}}^N\mathbb{I}_{m},
\end{eqnarray}
where $\mathbb{U}_n=\exp{(-i\delta tH_n / \hbar)}$ in the space $\mathcal{H}_S\otimes\mathcal{H}_{E_n}$, and $H_n$ is the total Hamiltonian of the system, the environment and their interactions in the $n^{th}$ interval. Here, $\mathbb{I}_m$ is the identity operator defined in the environment Hilbert space. A collective evolution of the system-environment combination, up to a time $n\delta t$ $(1\leq n\leq N)$ is given by $\rho\mapsto\overline{\mathbb{U}}_n\rho\overline{\mathbb{U}}_n^\dagger$, where the sequence of unitaries, $\{\overline{\mathbb{U}}_n\}$, satisfies 
\begin{eqnarray}
 \overline{\mathbb{U}}_{n+1}=\tilde{\mathbb{U}}_{n+1}\overline{\mathbb{U}}_n;\,\,\overline{\mathbb{U}}_0=\mathbf{\mathbb{I}},
 \label{eq:unitary}
\end{eqnarray}
with $\mathbf{\mathbb{I}}$ being the identity operator in  $\mathcal{H}_{tot}$. We will consider the unitary evolution given in Eq. (\ref{eq:unitary}) up to a time $N\delta t$,  in the limit $N\rightarrow\infty$ and $\delta t\rightarrow 0$, such that $N\delta t$ remains finite.

Let us now assume that at the beginning of the $n^{th}$ time interval of duration $\delta t$, the states of $S$ and $E_n$ are $\rho_S$ and $\rho_{E_n}$ respectively. Let us also assume that $\{B_n^j\}$ is the linearly independent basis on the operator space of $\mathcal{H}_{E_n}$, which are orthonormal with respect to the inner product $\langle A^1, A^2 \rangle_{\rho_{E_n}} = \mbox{tr}(\rho_{E_n} {A^{1}}^{\dagger} A^2)$, implying $\mbox{tr}(\rho_{E_n}{B_n^i}^{\dagger}B_n^j) =  \delta_{ij}$. Therefore,
\begin{eqnarray}
\mathbb{U}_n = \sum_j \mathbb{U}_n^j \otimes B_n^j,
\label{eq:L_form}
\end{eqnarray}
where $\{\mathbb{U}_n^j\}$ are operators on $\mathcal{H}_S$.

After the $n^{th}$ time interval, the state of $S$ evolves from $\rho_S$ to $\mathcal{D}_n(\rho_S)$, with
\begin{eqnarray}
\mathcal{D}_n(\rho_S) &=& \mbox{tr}_{E_n} \left(\mathbb{U}_n \rho_S \otimes \rho_{E_n} \mathbb{U}_n^{\dagger} \right) \nonumber \\
&=& \sum_{ij} \mathbb{U}_n^i \rho_S {\mathbb{U}_n^j}^{\dagger} \ \mbox{tr} (B_n^i \rho_{E_n} {B_n^j}^{\dagger}) \nonumber \\
&=& \sum_{j} \mathbb{U}_n^j \rho_S {\mathbb{U}_n^j}^{\dagger},
\label{eq:L_rho}
\end{eqnarray}
so that the quantum master equation corresponding to the $n^{th}$ interaction can be derived from
\begin{eqnarray}
\frac{d \rho_S}{d t}= \underset{\delta t \rightarrow 0}{\mbox{lim}} \frac{\mathcal{D}_n(\rho_S) - \rho_S}{\delta t}.
\label{eq:rho_der}
\end{eqnarray}
Noticing that all the spins in the collection are identical with $H_{E_i}\equiv H_E=B\sigma^z_E$ and $\rho_{E_i}\equiv\rho_E$,  Eqs. (\ref{eq:L_form}) and (\ref{eq:L_rho}) hold true for every interval, implying that discarding the index ``$n$",  Eq. (\ref{eq:rho_der}) provides the master equation for the entire evolution.

Now, the total system-environment Hamiltonian, $H$, given by Eqs. (\ref{eq:h_tot}) and (\ref{eq:h_int}), can be written as
\begin{eqnarray}
H = 
\begin{pmatrix}
H_S + B \mathbb{I}_S && 2\sqrt{k/\delta t}\sigma_d^- \\
2\sqrt{k/\delta t}\sigma_d^+ && H_S - B \mathbb{I}_S
 \end{pmatrix},
\end{eqnarray}
where $\sigma_d^{\pm}=\sigma_d^x\pm i\sigma_d^{y}$.  In turn, $\mathbb{U} = \exp(-i \delta t H / \hbar)$ can be written as
\begin{widetext}
\begin{eqnarray}
\mathbb{U} = 
\begin{pmatrix}
\mathbb{I}_S - \frac{\delta t}{\hbar} \big(i B \mathbb{I}_S + i H_S + \frac{2 k}{\hbar} \sigma_d^- \sigma_d^+ \big) + o(\delta t^2) && - \frac{2i}{\hbar} \sqrt{k \delta t} \sigma_d^- + o(\delta t^{3/2}) \\
- \frac{2i}{\hbar} \sqrt{k \delta t} \sigma_d^+ + o(\delta t^{3/2}) && \mathbb{I}_S + \frac{\delta t}{\hbar} \big(i B \mathbb{I}_S - i H_S - \frac{2 k}{\hbar} \sigma_d^+ \sigma_d^- \big) + o(\delta t^2)
\end{pmatrix}.
\end{eqnarray} 
\end{widetext}

We consider the thermal state $\rho_E=\mbox{diag}\{p_0,p_1\}$ of the environment at temperature $T_E$ to be its initial state, where 
\begin{eqnarray}
p_0 = Z_E^{-1}\exp(-\beta_E B),\, p_1= Z_E^{-1}\exp(\beta_E B),
\end{eqnarray}
with $Z_E^{-1}=\mbox{tr}\big[\exp(-\beta_E B \sigma^z_E)\big]$, and $\beta_E=(k_B T_E)^{-1}$, $k_B$ being the Boltzmann constant. From $\rho_E$, $\{B^j\}$ matrices can be defined as
\begin{widetext}
\begin{eqnarray}
B^0 = \mathbb{I}_E,\, 
B^1 = \frac{1}{\sqrt{p_0}}\begin{pmatrix}0 && 0 \\ 1 && 0 \end{pmatrix}, \,\,
B^2 = \frac{1}{\sqrt{p_1}} \begin{pmatrix}0 && 1 \\ 0 && 0\end{pmatrix},\,\,
B^3 = \frac{1}{\sqrt{p_0 p_1}}
\begin{pmatrix}
p_1 && 0 \\ 0 && -p_0
\end{pmatrix},
\end{eqnarray}
\end{widetext}
such that $\mbox{tr}(\rho_{\beta_E}{B_n^i}^{\dagger}B_n^j) =  \delta_{ij}$. The elements of $\mathbb{U}$ in the basis $\{B^j\}$ are given by
\begin{widetext}
\begin{eqnarray*}
\mathbb{U}^0 &=& \mathbb{I}_S + \frac{\delta t}{\hbar} \big( -i H_S + i B (p_1 -p_0)\mathbb{I}_S - \frac{2 k}{\hbar} p_0\sigma_d^-\sigma_d^+
- \frac{2 k}{\hbar} p_1\sigma_d^+\sigma_d^-\big)+ o(\delta t^2),  \nonumber \\
\mathbb{U}^1 &=& - \frac{2 i}{\hbar} \sqrt{p_0 k \delta t} \sigma_d^+ + o(\delta t^{3/2}), \,\,
\mathbb{U}^2 = - \frac{2 i}{\hbar} \sqrt{p_1 k \delta t} \sigma_d^- + o(\delta t^{3/2}), \,\,
\mathbb{U}^3 = o(\delta t).
\end{eqnarray*}
\end{widetext}
Using these, straightforward algebra leads to
\begin{widetext}
\begin{eqnarray}
\sum_j \mathbb{U}^j \rho_S {\mathbb{U}^j}^{\dagger} = \rho_S - \frac{i \delta t}{\hbar} [H_S, \rho_S] 
+ \frac{2 k p_0 \delta t}{\hbar^2} (2\sigma_d^+ \rho_S \sigma_d^- - \{\sigma_d^-\sigma_d^+, \rho_S\}) 
+ \frac{2 k p_1\delta t}{\hbar^2} (2\sigma_d^- \rho_S \sigma_d^+ - \{\sigma_d^+\sigma_d^-, \rho_S\})
+ o(\delta t^2).\nonumber \\
\end{eqnarray}
\end{widetext}
We retain terms upto $\delta t$, and obtain, from Eq. (\ref{eq:rho_der}), the Lindblad master equation given by Eq. (\ref{eq:qme}),  corresponding to local repetitive  interaction with a single door, $d$.  Redefining $\sigma_d^{\pm}$ as $\eta_{d}^\alpha=\sigma_{d}^x+i(-1)^\alpha\sigma_{d}^y$, the dynamical term is given by Eq. (\ref{eq:dynamical_term}) (cf. \cite{dhahri08}). Note here that Eq. (\ref{eq:dynamical_term}) describes a dissipation process with rate $\frac{4 k p_1}{\hbar^2} $ and an absorption process with rate $\frac{4 k p_0}{\hbar^2}$. For high values of $\beta_E$, $p_0 \approx 0$ and $p_1 \approx 1$, and the resulting dynamics is that of a Markovian amplitude-damping noise  \cite{petruccione}.


\begin{thebibliography}{200}
\bibitem{defLN} The logarithmic negativity (LN) \cite{vidal02} of a quantum state $\rho_{AB}$ is given by $\mathcal{L}(\rho_{AB}) = \log_2 \|\rho_{AB}^{T_A}\|_1$, where \(\rho_{AB}^{T_{A}}\) is the partially transposed form of $\rho_{AB}$ \cite{peres96} w.r.t $A$, and $\|\rho\|_1 \equiv \mbox{tr}\sqrt{\rho^\dag \rho}$.
\bibitem{dimertoPMII} Note that in the earlier papers \cite{divakaran16, dutta15}, the PM-II phase in the ATXY model was mentioned as the dimer phase. Our recent analysis \cite{dm_16} shows that the dimer order parameter \cite{dm_order} in this phase vanishes and it is indeed paramagnetic in nature. However, such finding does not affect the results obtained in this paper as well as the other results in \cite{divakaran16, dutta15}.
\end{thebibliography}
\end{document}